\documentstyle[11pt,newpasp,epsf,twoside]{article}
\def\edcomment#1{\iffalse\marginpar{\raggedright\sl#1\/}\else\relax\fi}
\reversemarginpar

\begin{document}
\title{Chemical Evolution in Hierarchical Clustering Scenarios}
 \author{Patricia B. Tissera}
\affil{Institute of Astronomy and Space Physics, CONICET, Argentina}
\author{
 Cecilia Scannapieco 
}
\affil{Institute of Astronomy and Space Physics, CONICET, Argentina}

\begin{abstract}
We present first results of an implementation of chemical evolution
in a cosmological hydrodynamical code,
 focusing the analysis on the effects of cooling baryons
according to their metallicity. We found that simulations with primordial
cooling can underestimate the star formation rate from $z < 3$ and 
 by up to $\approx 20 \%$. We constructed simulated spectra by combining
the star formation and chemical history of galactic systems with
spectra synthesis models and assess the impact of chemical evolution
on the energy distribution.

\end{abstract}

\section{Introduction}

The current cosmological paradigm  predicts a 
hierachical building up of the structure. This scenario has largely supported
by observations in the local and high redshift Universe.
However, few but important descrinpances remain to be explanied
such as the inner distribution of dark matter in small haloes,
the origin of the high baryonic content and 
metallicity of the intercluster medium, etc (Shanks 2004). Most of these
open problems requires the comprehensive understanding of galaxy formation,
including all the involved scales. 
Numerical simulations have played an important role in understanding and
proving the connection between theory and observations at different
scales. However, the modelling of small scale physical processes such
as gas cooling, star formation and chemical evolution within
a cosmological framework presents different difficulties because of 
our poor understanding of the relevant physics
and numerical limitations. In particular, only recently chemical
evolution has started to be treated within cosmological 
hydrodynamical simulations (Yepes et al. 1998; Cen \& Ostriker 1999; 
Mosconi et al. 2001).
  
The dramatic building up of observational evidences on chemical properties of
the stellar population and interstellar medium 
in the Universe has proved to
have a significant  impact on the study of galaxy formation.
Chemical elements are produced in the stellar interior and ejected in
a pacific or violent way after characteristic timescales determined by
stellar evolution. Afterwards,  these elements can  be   mixed up by 
  mergers and interactions. It is also important to
notice that the  metal production is directly linked to the star formation
history which can be also affected by the  environment (Barton et
al. 2000;
Lambas et al. 2003; Balogh et al. 2004). As a consequence, 
it is expected that  chemical patterns store information of the
process of galaxy formation ( Freeman  \& Bland-Hawthorn 2002).
 
Another important motivation to include a chemical treatment in galaxy
formation models is related to  the fact that
baryons cool according to their metallicity. Sutherland \& Dopita (1993)
showed that a gas cloud with solar metallicity cools at
a rate of up to few  orders of magnitude 
more efficiently than a primordial gas cloud.
Hence, the enrichment of the interstellar medium can have
dynamical  consequences and affect the formation of the structure.

In this work, we present first results of detail treatement  of
chemical evolution in Gadget2 (Springel \& Hernquist 2002) based
on the implementation of Mosconi et al. (2001).
In Section 2 we briefly summarize the main important aspects of the
chemical code. Section 3 shows the first results and Section 4
presents the conclusions.

\section{The model}

The chemical model has been implemented  within 
 Gadget2 (Springel \& Hernquist
2002) which is
based on a  modified version of SPH 
that conserves entropy (energy and angular momentum).
Stars are formed according to a probability function defined by
the relation between cooling and dynamical time (see also Lia et al. 2001).
 A  standard Salpeter Initial mass function has been assumed.
Supernovae 
II are considered to form from stars with $M > 8 M_{\odot}$
and to produce metals according to Woosley \& Weaver (1995) yields.
Supernova Ia are assumed to  originate from binary systems following
the  model of Thielemann et al. (1993). We have assumed a random
sampling of the lifetime of binary systems in the range 0.1-5 Gyrs.

Chemical elements are distributed within gas neighbouring particles by using a
spline kernel function.
In order to estimate the correct environment of stars when SN explode we
re-calculate the gaseous neighbours of the stellar
 particles each time elements
are released.
In this implementation we have not included the treatment of energy feedback.

 
We  run cosmological tests of $\Lambda$-CDM cosmologies 
 consistent with the concordance
model ($\Omega_{\Lambda}=0.7$, $\Omega_{m}$=0.3, $H_o =70 \ {\rm km
s^{-1} Mpc^{-1}}$).
 The simulations correspond to a volume box of $10 h^{-1}$ Mpc 
($h=0.7$)
with $2 \times 50^3$ particles. Three different runs
were performed using different cooling functions: solar, primordial and
metal-dependent.

\section{Results}

The cosmological simulations are analyzed from a global point
of view and then, by identifying the gravitational bounded structures. 
In Figure 1 (left panel) 
we show the comoving star formation $\rho_{\rm SFR}$ and
metal production $\rho_{\rm Z}$ 
rates for the three runs with solar (solid lines),
primordial (dashed lines) and metal-dependent (dotted-dashed lines)
cooling functions.
 We can see that the largest
difference is found for the run with fixed solar cooling  and for 
$z \le 3$. 
The $\rho_{\rm SFR}$ of the metal-dependent cooling run
 can be up to  $\approx 50\%$ lower than that of the solar cooling test,
and up to $\approx 20\%$
 higher than the primordial  cooling run.
Similar relations  are found for the metal production
rate $\rho_{\rm Z}$ .
 Figure 1 (right panel)  shows the
fraction of gas with $T > 10^{5}$ K  for the three runs,
 with 
the largest
gas fraction associated to the primordial cooling run, as expected.
the difference between primordial and metal-dependent cooling are
localized in the densest regions.

At $z=0$ we identified galaxy-like objects (GLOs) at their virial radius by
combining the friends-of-friends algorithm with a density contrast method
in the three simulations (with solar, primordial and metal dependent cooling).
 We constructed the star formation history
of the  GLOs and analyzed the chemical properties of their stellar population
and interstellar medium.
In Figure 2a we show the SFR history of a typical GLO 
in the three runs. In the small box we have plotted the ratio
of the SFR of the solar and metal-dependent cooling to that of the 
primordial one.
We can see that the differences are important for $z <3$,
but do not present a regular behavior.
 This suggests that 
the non-linear evolution of the structure
determines  complicate star formation history and chemical patterns that can produce
non-linear responses to the characteristics of the cooling functions. Our model
will allow us to try to understand the link between
the dynamical evolution and the  star formation history and chemical patterns,
by studying the particular history of evolution of
GLOs in different environments.
 
\begin{figure}[t]
\plottwo{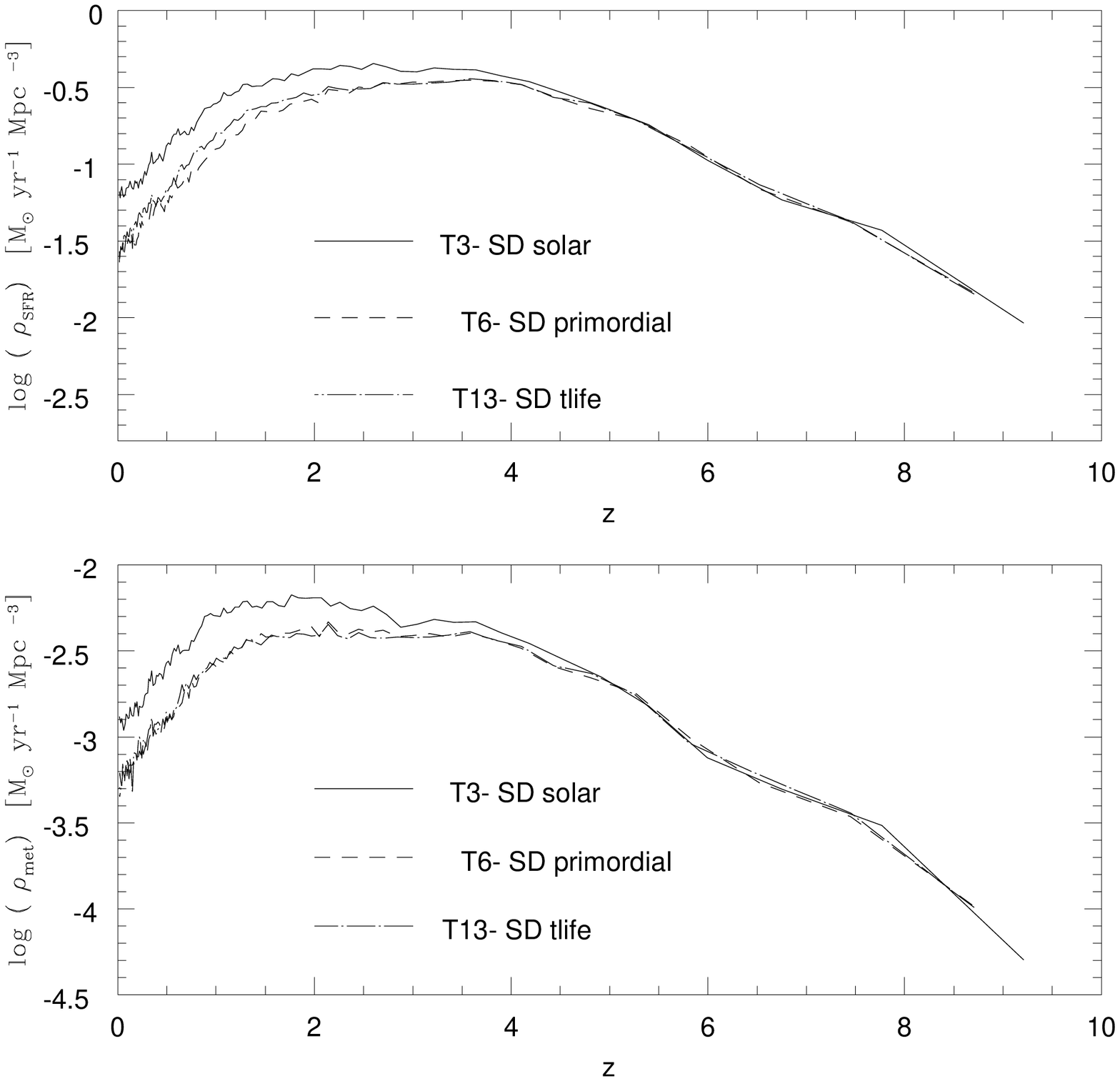}{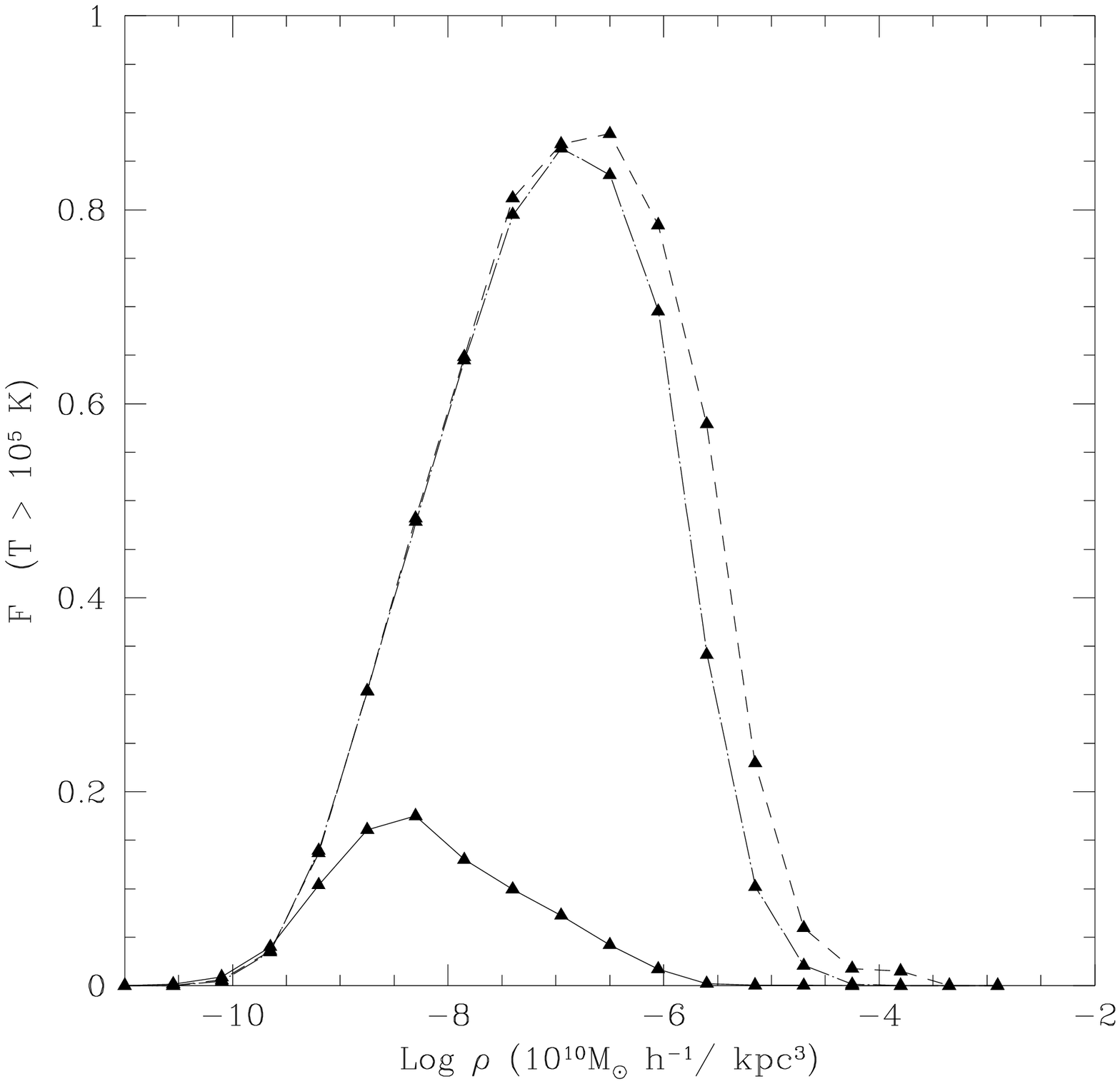} 
\caption{Comoving star formation and metal production rate for
three different cosmological runs of the same initial condition, where the
gas has been allowed to cool according to solar  (solid lines),
primordial (dashed lines)
and metal dependent (dotted-dashed lines) cooling functions.
 As it can be seen in the right 
panels, the largest differences are found for $z <3$. We
found at most a difference of up to $\approx 20 \%$ between the comoving
star formation rates of primordial and metal
dependent cooling runs.
}
 \label{rates}            
\end{figure}

\begin{figure}[t]
\plottwo{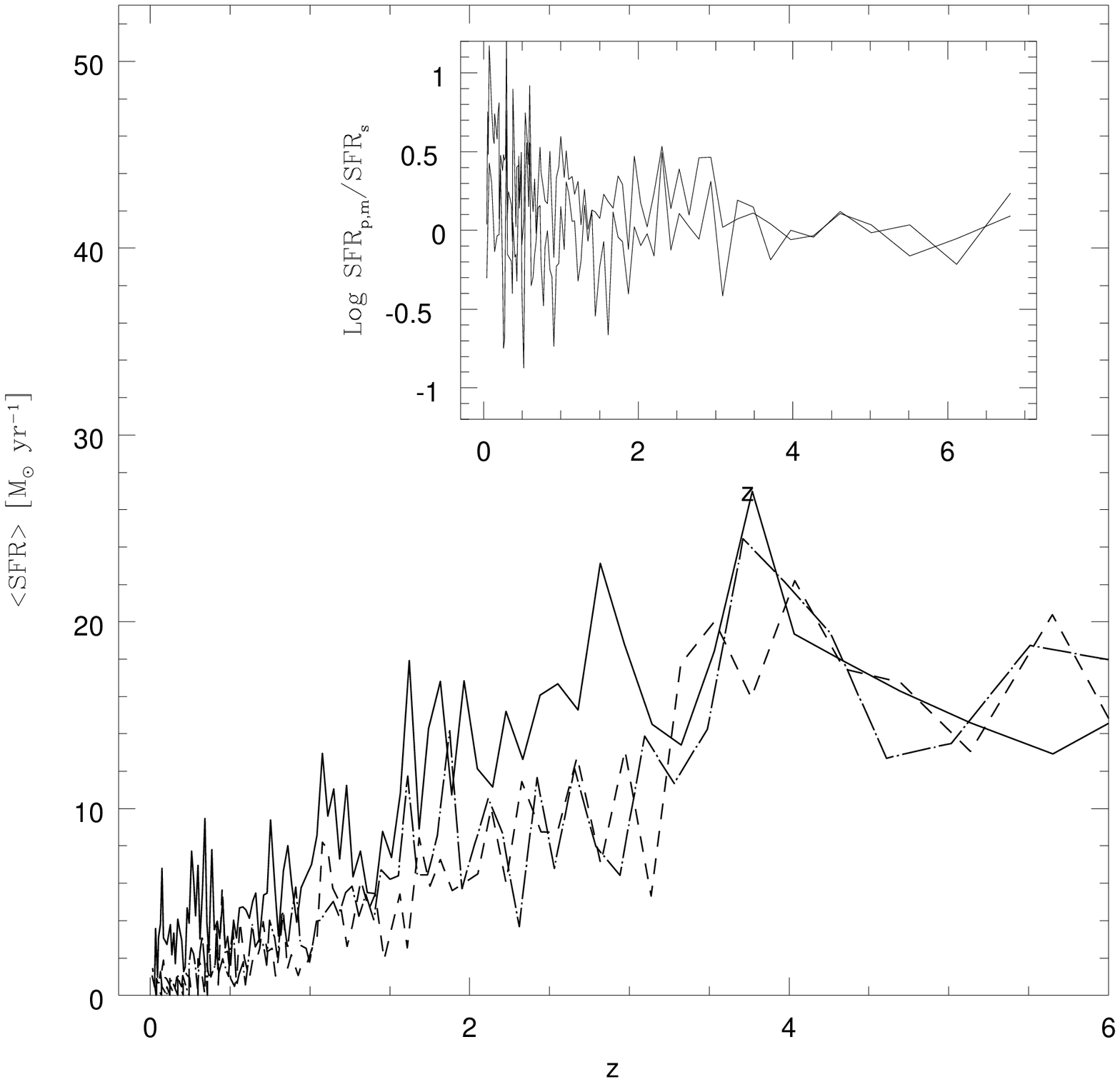}{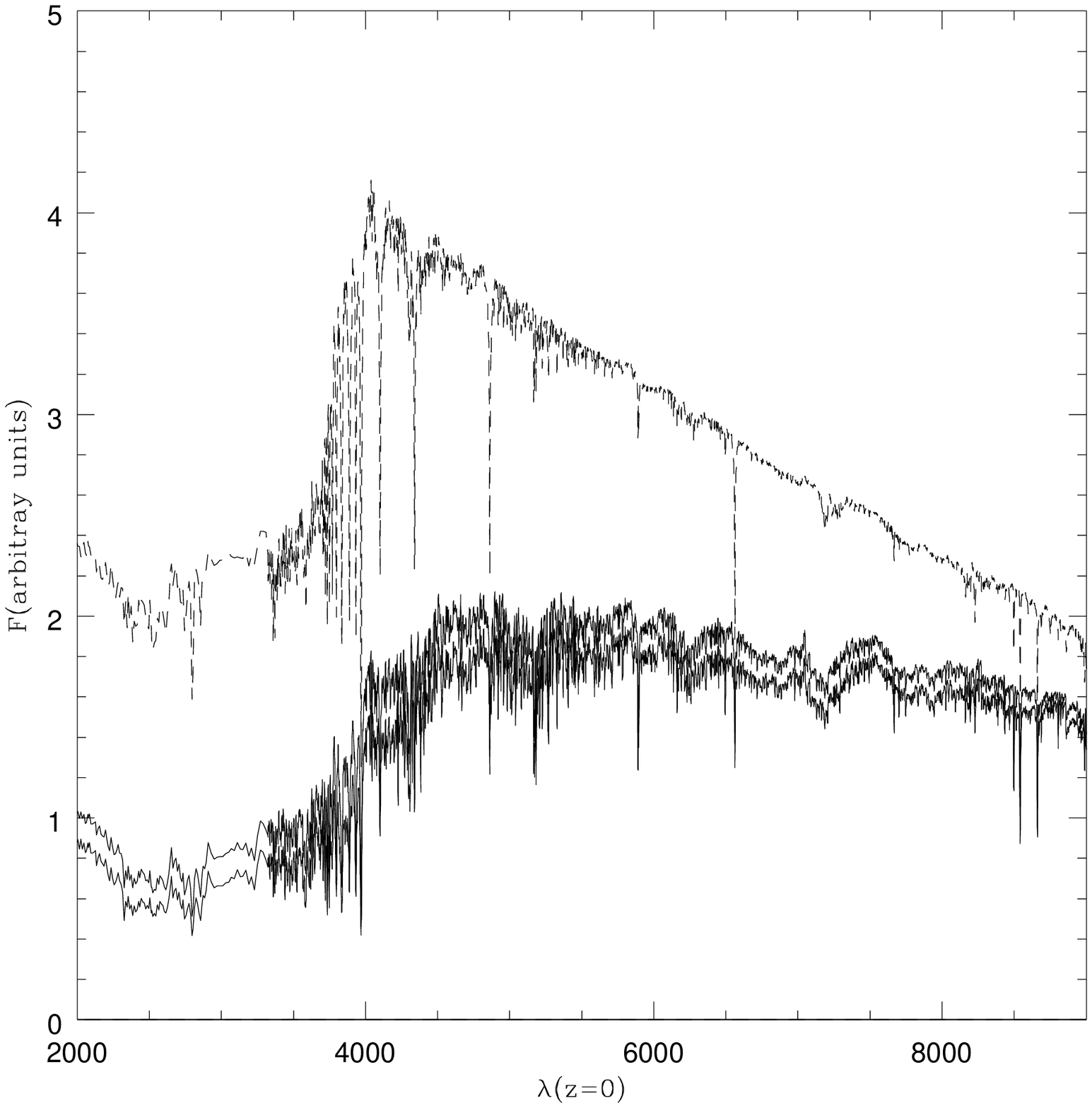} 
\caption{Letf panel: Star formation rate history 
for a typical galaxy-like object  
three different cosmological runs of the same initial condition, where the
gas has been allowed to cool according to solar  (blue solid lines),
primordial ($Z=0.0001 Z_{\odot}$, green dashed lines)
and metal dependent (red dotted-dashed lines) cooling functions.
Right panel: Simulated spectra by using the star formation rate history
of the metal dependent cooling run and the SED of BC03 for
$Z=0.0001 Z_{\odot}$ (green lines), $Z=Z_{\odot}$ (blue lines)
and for the corresponding metallicities of the stars (red lines).
}
 \label{SED-glo1}            
\end{figure}

It was shown by Tissera, Lambas \& Abadi (1997) that the combination
of the history of star formation with models of population synthesis 
 models results in a powerful tool to cnonect theory with
observations. The treatment of the chemical enrichment in the simulations
improve the technique since each simulated stellar population has its age and
metallicity estimated consistently with the history of formation
in a hierarchical scenario.
With this new chemical code, we can also assess the effects of cooling 
 the gas according to its metallicity.
For this purpose, we constructed the synthetic spectral distribution 
of the GLOs but combining the SEDs of Bruzual \& Charlot (2003, BC03) 
according to the metallicity and age of stars.
We also estimated the simulated spectra (SS) of the GLOs by assuming
fixed metallicity at $Z=0.0001 Z_{\odot}$ and $Z=Z_{\odot}$ for the SEDs
of BC03.
In Figure 2b we show the SS for the GLOS of 
~\ref{SED-glo1}a for the three runs.
Note that the SS has been calculated with the same SFR history
(~\ref{SED-glo1}a, solid line).
From this figure, we can see that assuming a low fixed  metallicity leads
to a very different energy distribution. For this GLO, the real SS is
more similar to that produced by assuming fixed solar metallicity.

We also analyzed the simulated spectra for the same GLO in the three
runs (solar, primordial and metal dependent cooling), and estimated
the SS by using the full information on age and metallicity.
Hence, in this case we will looking at the impact of different gas cooling
on the star formation and chemical history through the
differences in  the energy distribution.
Figure 3 we show
the evolution of the SS for the  GLO of Figure 2  in the three runs.
As it can be seen, the GLOs in the solar cooling run exhibits the
more luminous spectra since because of the high efficient cooling, 
a large fraction of stars formed quickly before the gas has time to enriched.
In the case of the corresponding GLO in the runs with primoridal and
metal dependent cooling, the interstellar medium has time to get
enriched by previous star generations. Hence, in this case, the 
SS of the GLO metal-dependent cooling run is more similar to that
obtained from the primordial one.

\begin{figure}[t]
\plotfiddle{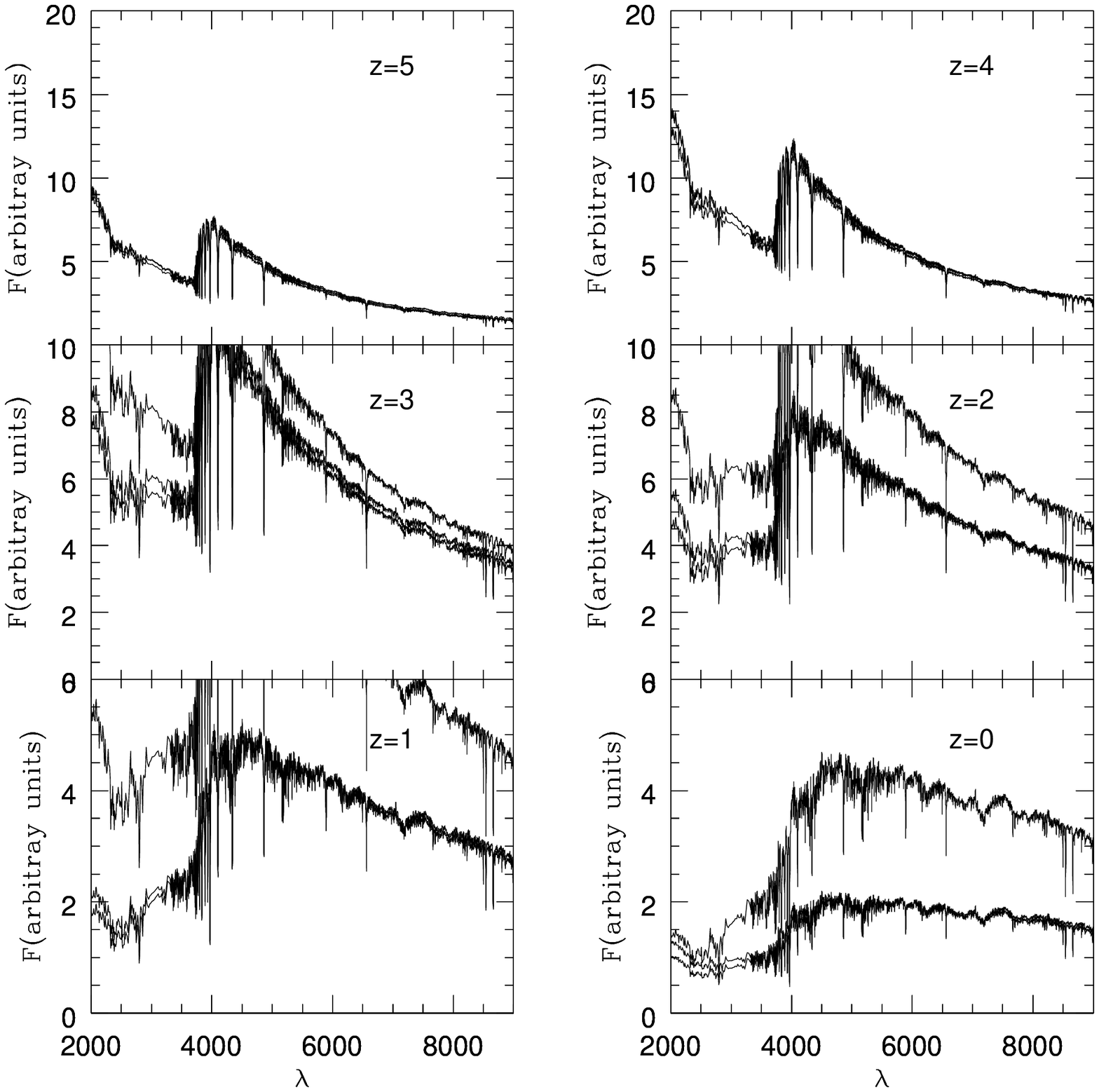}{2.4in}{0}{50}{40}{-160}{-85}
\caption{Simulated spectra of the GLO shown in Figure 2   as a function
of redshift for  the 
three different cosmological runs of the same initial condition, where the
gas has been allowed to cool according to solar  (blue solid lines),
primordial (green dashed lines)
and metal dependent (red dotted-dashed lines) cooling functions.
}
 \label{CF-glo1-evol}            
\end{figure}

\section{Conclusions}

We have incorporated a chemical evolution algorithm within
Gadget2 which allows the description of the enrichment of baryons
within a cosmological framework.  A first analysis shows
that in $\Lambda$-CDM scenarios,  the comoving star formation rate
could be underestimated by up to $ \approx 20\%$  
if the  primordial cooling function is adopted.
However, within individual galactic systems it is complicated
to disentangle 
the effects of cooling the gas consistently with
its metallicity owing to the non-linear growth of the structure
which affects the star formation,  chemical histories and the mixing of metals
in a non-linear way.
 First results show that the star formation history
 could be either
underestimated or overestimated by an order of magnitude depending
on the particular history path.
We also show the simulated spectra obtained by combining the age
and metallicity information of stars in the simulation with
the SEDs of Bruzual \& Charlot (2003).Future works will focus
on the study of  the link between the characteristic of the spectra and
history of evolution of the galactic systems.



\section*{Acknowledgments}

We thank Simon White and Volker Springel for useful discussions,
and V.Springel for making Gadget2 available for this work.
We acknowledge the support of Fundanci\'on Antorchas
and  DAAD that made this work possible. 
C. Scannapieco thanks the Alexander von Humboldt Foundation, the Federal Ministry
of Education and Research, and the Programme for Investment in the Future (ZIP)
of the German Government for partial support.
This work was partially funded by CONICET.

\end{document}